\documentclass[dvipdfmx, 12pt] {article}
\usepackage{tikz}
\usepackage{pgf}
\usetikzlibrary{arrows}
\usepackage{here}

\usepackage{amsmath, amsthm, amsfonts, amssymb, color}
\usepackage{color}
\usepackage{ascmac}

\newcommand{\beq}{\begin{eqnarray*}}
\newcommand{\eeq}{\end{eqnarray*}}
\newcommand{\bet}{\begin{tikzpicture}}
\newcommand{\ent}{\end{tikzpicture}}

\newsavebox{\toy}
\savebox{\toy}{\framebox[0.65em]{\rule{0cm}{1ex}}}
\newcommand{\QED}{\usebox{\toy}}
\def\nlni{\par\ifvmode\removelastskip\fi\vskip\baselineskip\noindent}

\begin{document}

%%%%%%% DOUBLE SPACED %%%%%%%%
\setlength{\baselineskip}{15pt}
\title{
Limiting 
distribution of extremal eigenvalues of d-dimensional random Schr\"odinger operator
}
\author{
Kaito Kawaai, 
Yugo Maruyama,
and
Fumihiko Nakano
\thanks{
Mathematical Institute,
Tohoku University,
Sendai 980-8578, Japan
%
%e-mail : 
%fumihiko.nakano.e4@tohoku.ac.jp
}
}
%\date{最終更新日：}
\maketitle
%第一ページの番号を消す
%\thispagestyle{empty}
%%%%%%% ABSTRACT %%%%%%%%%%%%%
\begin{abstract}
We consider 
Schr\"odinger operator with random decaying potential on 
$\ell^2 ({\bf Z}^d)$ 
and showed that, 
(i)
IDS 
coincides with that of free Laplacian in general cases, 
(ii) 
the set of extremal eigenvalues, 
after rescaling, 
converges to a inhomogeneous Poisson process, 
under certain condition on the single-site distribution, 
and 
(iii)
there are ``border-line" cases, 
such that
we have Poisson statistics in the sense of (ii) above if the potential does not decay, 
while 
we do not if the potential does decay. 

\end{abstract}

%Mathematics Subject Classification (2000): 82B44, 81Q10

%\tableofcontents
%%%%% INTRODUCTION %%%%%%%%%%%%%%%%%%%%%%%%%%%%%%%%%
\section{Introduction}
The study 
on Schr\"odinger operators with random decaying potential on 
$\ell^2 ({\bf Z}^d)$ 
was initiated by 
\cite{DKS}, 
where 
they showed that if 
$d=1$, 
it has various spectral properties depending on the decay rate of the potential.
And 
a decade later, 
spectral properties related to Anderson localization were  studied for general 
$d$-dimension 
\cite{KKO}.
Recently, 
topics related to eigenvalue/eigenfunction statitics are studied for 
$d=1$ : e.g., 
eigenvalue statistics for the bulk
\cite{KVV, KN1, N1, KN2},
linear statistics 
\cite{N2, B},
eigenfunction statitics
\cite{RV, N3},
and limiting distribution of the maximal eigenvalue
\cite{D}. 
In this paper, 
we consider the Schr\"odinger operator with random decaying potential for general 
$d$-dimension, as is studied by \cite{D} : 
\beq
H
&:=&
H_0 + V, 
\quad
%\omega \in \Omega
\\
(H_0 u)(n)
&=&
\sum_{|m-n|=1}
u(m),
\\
(V u)(n)
&:=&
\frac {
\omega_n}
{
\langle n \rangle^{\alpha}
}
u(n), 
\quad
\langle n \rangle :=
(1 + |n|), 
\quad
\alpha \ge 0, 
\quad
u \in \ell^2({\bf Z}^d).
\eeq
where 
$\{ \omega_n \}_{n \in {\bf Z}^d }$ : 
is i.i.d. 
with common distribution 
$\mu$. 
Let 
$L \in {\bf N}$
and we set the box 
$\Lambda_L$ 
of size 
$2L+1$ 
and the finite-box Hamiltonian 
$H_L$
which is the restriction of 
$H$ 
on
$\Lambda_L$ : 
\beq
\Lambda_L
&:=&
\left\{ 
n=(n_1, \cdots, n_d) \in {\bf Z}^d
\, \middle| \,
|n_i| \le L, 
\;
i = 1, 2, \cdots, d
\right\}
\\
H_L
&:=&
1_L H 1_L, 
\quad
(1_L)(n) := 1(n\in\Lambda_L).
\eeq
Dolai \cite{D}
showed that, when 
$\mu$
has finite second moment : 
${\bf E}[ \omega_0^2 ] < \infty$, 
then the 
IDS
of 
$H$ 
is equal to that of the free Laplacian. 
Moreover, 
if the tail of 
$\mu$ 
satisfies 
$\mu[ x, \infty) = x^{-p}$, 
$p > 0$, 
he obtained the limit distribution of the maximal eigenvalue of 
$H_L$.
The purpose 
of this paper is to extend his result to show : 
(1)
the same conclusion for IDS is valid without any restriction on 
$\mu$, 
%and 
(2)
the joint distribution of the rescaled extremal eigenvalues converges to a Poisson process 
under more general condition on the tail of 
$\mu$, 
and
(3)
if the tail of 
$\mu$
is exponentially decaying, we have the same conclusion as in (2) if 
$\alpha = 0$, 
but do not if 
$\alpha > 0$.
\\
%

%
%(1)
To set up the problem, let 
$E_j (L)$, 
$j = 1, 2, \cdots, |\Lambda_L|$
be the eigenvalues of 
$H_L$ 
and let
$\mu_L$ 
be the empirical measure for the eigenvalues of 
$H_L$ 
: a random probability measure on 
${\bf R}$
defined by 
\begin{equation}
\mu_L
:=
\frac {1}{| \Lambda_L |}
\sum_j
\delta_{ E_j (L) }.
\label{empirical}
%\quad\cdots (*)
\end{equation}
Among 
many known results, we recall 
(i)
let 
$\mu_L^0$
the empirical measure for the free Laplacian 
(that is, the Hamiltonian 
$H_0$).
Then 
we have an ac probability measure 
$\mu^0$
with 
supp $\mu^0 = [-2d, 2d]$ 
such that 
$\mu_L^0 
\stackrel{w}{\to}
\mu^0$.
(ii)
%\cite{D}
if 
${\bf E}[ \omega_0^2 ] < \infty$, 
we have 
$\mu_L 
\stackrel{w}{\to}
\mu^0$
(\cite{D}).
We first 
remark that the second moment condition in 
\cite{D} 
is not necessary : \\
%

%
%%%%%
{\bf Theorem 1 }
{\it 
Let 
$\alpha > 0$. 
For any i.i.d.
$\{ \omega_n \}$, 
we have }
\beq
\mu_L^{\omega} 
\stackrel{w}{\to}
\mu^0, 
\quad
a.s.
\eeq
%

%%%%%
%
We turn 
to study the extremal eigenvalues.
We denote 
the tail of common distribution 
$\mu$ 
by 
\beq
\mu [ x, \infty )
=
\frac {1}{f(x)}, 
\quad
x > 0, 
\eeq
for a function 
$f$.
Let 
$\{ E^H_j (L) \}_{j \ge 1}$ : 
$E^H_1(L) \ge E^H_2(L) \ge \cdots$ 
be positive eigenvalues of 
$H_L$ 
in decreasing order, and let 
\beq
\widetilde{E}^{H}_j
:=
\frac {
f(E^{H}_j (L))
}
{
\Gamma_L
}, 
\quad
j = 1, 2, \cdots, 
\eeq
be the scaling of those, where 
$\Gamma_L$ 
will be chosen depending on 
$f$ 
such that 
$\lim_{L \to \infty}
\Gamma_L = \infty$. 
We set 
the following two assumptions on 
$f$ 
and 
$\Gamma_L$.
\\
%

%

%
%%%%%%%%%
{\bf Assumption 1 }
{\it 
$f : (0, \infty) \to (0, \infty)$
and 
$\Gamma_L$
satisfy the following conditions : \\
(1)
$f$
is strictly increasing on 
$[R, \infty)$ 
for some 
$R > 0$, 
$\lim_{x \to \infty} f(x) = \infty$, 
$f \in C^1$, 
and 
$\lim_{L \to \infty}
\Gamma_L = \infty$, 
\\
(2)
$f'(x) = o\left( f(x) \right)$, 
$x \to \infty$
\\
(3)
$\sup_{|x-y| \le 2d}
|f(y)|
\le
C
|f(x)|$
for sufficiently large 
$x$ 
and a positive constant 
$C$. 
}
\\

The condition 
(1) 
is natural, since  
$1/f$ 
gives the tail of a measure.
Conditions 
(2), (3) 
are satisfied if 
$f$
is of regular variation.
On the other hand, 
the following one is essential for our problem and non-trivial : \\
{\bf Assumption 2 }
{\it 
Let 
\beq
p_n^{(L)} (x)
:=
{\bf P}
\left(
\frac {
f(V(n))
}
{
\Gamma_L
}
\ge x
\right), 
\quad
n \in \Lambda_L, 
\quad
x > 0
\eeq
then we have
} 
\beq
\lim_{L \to \infty}
\sum_{n \in \Lambda_L} 
p_n^{(L)} (x)
=
\frac 1x, 
\quad
x > 0.
\eeq
%
%%%
%

We note that, if 
$\alpha = 0$
and 
Assumption 1
is satisfied,  
Assumption 2 
is always valid with 
$\Gamma_L = | \Lambda_L |$.
In what follows, 
we omit the 
$L$-dependence of 
$p_n$
for simplicity.
Under the two assumptions above, 
the rescaled extremal eigenvalues converge to a Poisson process : \\
%

%
%%%%%%%%%
{\bf Theorem 2 }
{\it 
Suppose 
$f$, 
$\Gamma_L$ 
satisfy Assumption 1, 2, and let 
$\nu$ 
be the measure on 
$(0, \infty)$
with 
$\nu[ x, \infty ) = \dfrac 1x$.
Then 
the point process
$\xi_L
:=
\sum_{j \ge 1}
\delta_{ \widetilde{E}^H_j(L) }$
with atoms being composed of the rescaled eigenvalues converge in distribution to 
Poisson 
($\nu$), 
where 
Poisson 
($\nu$) 
is the Poisson process with intensity measure 
$\nu$.
}\\

Here 
we consider the vague topology on the space of point processes on 
${\bf R}$.
As for the related results, 
the eigenvalue statistics 
on the bulk for 
$d=1$ 
is well studied 
\cite{KVV, KN1, N1, KN2}
and the various limits such as clock, Sine$_{\beta}$ and Poisson appear. 
However, 
there are not so many papers on the behavior of extremal eigenvalues even for 
$\alpha= 0$. 
~~We have  
two classes of functions satisfying Assumption 1, 2. 
The first one 
%the following family of functions 
%$f_{p,q}$
is a family of power functions with some logarighmic corrections.
%which satisfy Assumption 1, 2.
%
%
\beq
f(x)
=
f_{p, k}(x)
&:=&
x^p 
(\log x)^{-k}, 
\quad
p > 0, 
\quad
%q \in {\bf R}, 
k \in {\bf N} \cup \{ 0 \}, 
\quad
d \ge \alpha p.
\eeq
%
%
%For 
%technical reasons, we have to assume 
%$q$ 
%is an nonpositive integer : 
%we set 
%$q = -k$, 
%$k \in {\bf N} \cup \{ 0 \}$. 
%
We remark that 
Dolai 
\cite{D} 
obtained the limiting distribution of 
$\widetilde{E}^H_1 (L)$ 
when 
$p > 0$, $k=0$. \\
%

%
%%%%%%%%%
{\bf Theorem 3 }
{\it 
$f_{p, q}$, 
$\Gamma_L$ 
satisfy Assumption 1, 2 
if \\
(1)
$\alpha p < d$ : 
%$q = k$ :  
%$k \in {\bf N} \cup \{ 0 \}$ :  : 
%
%
\beq
\Gamma_L
:=
\gamma_{p, k} 
L^{d - \alpha p}, 
\quad
\gamma_{p, k}
:=
\frac {
C_{d-1}
}
{
d - \alpha p
}
\left(
\frac {d}{d - \alpha p}
\right)^k
\eeq
(2)
$\alpha p = d$ : 
%$q = -k$, $k \in {\bf N} \cup \{ 0 \}$ : 
%
\beq
\Gamma_L
=
h_k^{-1}
\left(
\gamma_k
(\log L)^{k+1} 
\right), 
\quad
h_k(x) := x (\log x)^k, 
\quad
\gamma_k 
:=
\frac {C_{d-1}}{k+1}
\cdot
p^k
%\frac {1}{C_{d-1}}
%\frac {k+1}{p^k}
\eeq
where 
$C_{d-1}
:=
|S^{d-1}|$
is the surface area of the d-dimentional unit ball.\\
}
%%%

%
We believe that for 
$\alpha p < d$ 
the result is true for any 
$k \in {\bf R}$.
Note that 
Theorem 3(1)
includes the case for the usual Anderson model where  
$\alpha = 0$.
It is natural 
to expect that the statement in Theorem 3 would be valid for general function 
$f$
which is of regular variation of order 
$p$ : 
\begin{equation}
\lim_{x \to \infty}
\dfrac {f(\lambda x)}{f(x)}
=
\lambda^p, 
\quad
\lambda > 0.
\label{RV}
%\quad\cdots (\sharp)
\end{equation}
In fact, 
a formal computation indicates that 
$f$
would satisfy Assumption 2 with 
$\Gamma_L = L^{d - \alpha p}/(d - \alpha p)$
(say, for the case of 
$\alpha p < d$).
However, 
the constant 
$\gamma_{p, k}$
in Theorem 3 (1) 
implies that these observation is false in general and the quantity which vanishes in the limit in 
(\ref{RV}) 
has a non-zero contribution in the limiting behavior of 
$\xi_L$.

~~We next consider
a family of exponential functions : 
\beq
f(x)
&=&
f_{\delta}(x)
:=
e^{ x^{\delta} }, 
\quad
0 < \delta \le 1. 
\eeq
In this case, the tail of 
$\omega_n$
is smaller than the previous one, so that we expect the behavior of eigenvalues become more gentle. 
Here 
we modify the definition of the norm of 
$n \in {\bf Z}^d$ 
for the sake of simplicity : 
\beq
\langle n \rangle := 1 + |n|_{\infty}, 
\quad
|n|_{\infty}
:=
\max_{i=1, 2, \cdots, d} |n_i|.
\eeq
Since 
$|n|_{\infty}
\le
|n|
\le
\sqrt{d} |n|_{\infty}$, 
this is not an essential modification.
The following 
Theorem implies these exponential functions are on a borderline. \\

{\bf Theorem 4}\\
%A
{\it 
(1)
$0 < \delta < 1$, 
$\alpha = 0$ : 
$f_{\delta}$
satisfies Assumption 1,2 with 
$\Gamma_L = |\Lambda_L|$. 
\\
(2)
$0 < \delta \le 1$, 
$\alpha > 0$ : 
we can find 
positive constants 
$C_j$, 
$j=1,2$ 
such that for sufficiently large
$x$, 
we have 
\beq
1 - C_1 e^{-x^{\delta}}
\le
{\bf P}
\left(
\bigcap_{L \ge 1}
\left\{
E^H_1 (L) \le x
\right\}
\right)
%
%\liminf_{L \to \infty}
%{\bf P}
%\left(
%E^H_{1} \le x
%\right)
%\le
%\limsup_{L \to \infty}
%{\bf P}
%\left(
%E_{1}^H \le x
%\right)
\le
\exp
\left[
- C_2
x^{- d / \alpha}
e^{- 2 D_{\alpha, \delta} x^{\delta} }
\right]
\eeq
where 
$D_{\alpha, \delta}
=
\max \{1, 2^{\alpha \delta - 1} \}$.
}\\
In Theorem 4(2), 
the first inequality 
implies that the sequence 
$\{ E_1^H (L) \}_{L \ge 1}$
is bounded for probability close to 
$1$, 
while the second one implies that there is no constant 
$M$
such that 
$E_{1}^H (L)\le M$, a.s., 
having completely different behavior from that in Theorem 4(1). 

~~In later sections, 
we prove these theorems. 
The proof of 
Theorem 1 is simple : we cut off 
$\omega_n$ 
suitably and let 
$\widetilde{H}$
be the corresponding Hamiltonian. 
Since 
its single site distribution has finite second moment, the IDS of 
$\widetilde{H}$
converges to 
$\mu^0$ 
by the result of 
\cite{D}.
It then suffices 
to compare IDS' of 
$H$
and 
$\widetilde{H}$
by the rank inequality. 
For the proof of Theorem 2, 
we first consider the point process 
$\xi_L^V$
composed of the rescaled eigenvalues of the multiplication operator 
$V$
and show that it converges to 
Poisson 
($\nu$), 
which is a non-identically distributed version of Poisson limit theorem.
We then 
compare 
$\xi_L$
and 
$\xi^V_L$ 
by using Assumption 1 (2).
For the proof of Theorem 3, 
we examine the condition in Assumption 2 explicitly for given 
$f_{p,q}$
and 
$\Gamma_L$.
The proof 
of Theorem 4(2) is reduced to the study of 
$E^V_1(L)$
and we can explicitly compute the quantity in question. 
%
%%%%%%%%%%%%%%%%%%%%%%%%%%%%%%%%%%%%%%%%%%%%%%%%%%%%%%%%%%%%
\section{Proof of Theorem 1}
We introduce 
a cut off parameter
$K \gg 1$
and set 
\beq
\widetilde{\omega}_n
&:=&
\omega_n
1 \left( |\omega_n| \le K \right), 
\quad
(\widetilde{V} u)(n)
:=
\frac {
\widetilde{\omega}_n
}
{\langle n \rangle^{\alpha}}
u(n), 
\\
\widetilde{H}
&:=&
H_0 + \widetilde{V}.
\eeq
Let 
$\widetilde{\mu}_L$
be the empirical measure for 
$\widetilde{H}$.
Since 
$\{ 
\widetilde{\omega}_n \}$
is i.i.d. with 
${\bf E}[ \widetilde{\omega}_0^2 ] \le K^2 < \infty$, 
the result in 
\cite{D} 
says that 
$\widetilde{\mu}_L 
\stackrel{w}{\to}
\mu^0$
which is equivalent to 
\begin{equation}
\quad
d_L (\widetilde{\mu}_L, \mu^0)
\stackrel{L \to \infty}{\to} 0, 
\quad
a.s.
\label{one}
%\quad \cdots (1)
\end{equation}
where 
$d_L (\mu, \nu)$
is the L\'evy distance between the probability measures 
$\mu$
and 
$\nu$.
For any 
$\epsilon > 0$, 
we can find 
$K_{\epsilon} \gg 1$
sufficient large such that 
\beq
{\bf P}( 
| \omega_n | > K_{\epsilon}
)
< \epsilon.
\eeq
Set 
$K = K_{\epsilon}$ 
in the definition of 
$\widetilde{H}$, 
and let 
$d_{KS}(\mu, \nu)$
be the 
Kolmogorov—Smirnov distance between 
$\mu$
and 
$\nu$. 
By the rank inequality\cite{Bai}, 
\begin{equation}
d_{KS}
(\mu_L, \widetilde{\mu}_L)
\le
\frac {1}{| \Lambda_L |}
rank
(H_L - \widetilde{H}_L)
\le
\frac {1}{| \Lambda_L |}
\sum_{n \in \Lambda_L}
1(| \omega_n | > K_{\epsilon}).
\label{two}
%\quad \cdots (2)
\end{equation}
By the strong law of large numbers, for any 
$\epsilon > 0$, 
we can find 
$\Omega_{\epsilon}(\subset \Omega)$
with 
${\bf P}(\Omega_{\epsilon}) = 1$
such that 
\begin{equation}
\frac {1}{| \Lambda_L |}
\sum_{n \in \Lambda_L}
1(| \omega_n | >K_{\epsilon})
\to
P (| \omega_0 | > K_{\epsilon} )
< 
\epsilon,
\quad
\omega \in \Omega_{\epsilon}
\label{three}
%\quad \cdots (3)
\end{equation}
On the other hand, 
by the triangle inequality for 
$d_L$ 
and by 
$d_L(\mu, \nu) \le d_{KS}(\mu, \nu)$, 
\beq
d_L (\mu_L, \mu^0)
& \le &
d_L( \mu_L, \widetilde{\mu}_L)
+
d_L (\widetilde{\mu}_L, \mu^0)
\le
d_{KS}( \mu_L, \widetilde{\mu}_L)
+
d_{L} (\widetilde{\mu}_L, \mu^0).
\eeq
We take 
$\limsup_{L \to \infty}$
on both sides and use 
(\ref{one}, \ref{two}, \ref{three}).
%(1), (2), (3).
%
%
\beq
\limsup_{L \to \infty}
d_L (\mu_L, \mu^0)
\le
P (| \omega_0 | > K_{\epsilon} )
< 
\epsilon, 
\quad
\omega \in \Omega_{\epsilon}.
\eeq
Letting  
$\Omega_0 := \bigcap_{n \ge 1} \Omega_{1/n}$, 
we have 
${\bf P}(\Omega_0) = 1$
and
\beq
\lim_{L \to \infty}
d_L (\mu_L, \mu^0) = 0, 
\quad
\omega \in \Omega_0.
\eeq
\QED
%
%
%%%%%%%%%%%%%%%%%%%%%%%%%%%%%%%%%%%%%%%%%%%%%%%%%%%%%%%%%%%%
\section{Proof of Theorem 2}
We 
first prepare a lemma, which gives a sufficient condition to have a Poisson limit theorem, and follows easily from Assumption 1, 2.\\
%

%
%%%%%
{\bf Lemma 1 }
{\it 
Following equations 
are valid. 
\beq
&(0)& \quad
\lim_{L \to \infty}
\max_{n \in \Lambda_L}
p_n (x)
= 0, 
\quad
x > 0,
\\
&(1)&\quad
\lim_{L \to \infty}
\sum_{n \in \Lambda_L}
\frac {p_n(x)}{1 - p_n(y)}
= 
\frac 1x, 
\quad
x, y > 0,
\\
&(2)&\quad
\lim_{L \to \infty}
\sum_{n \in \Lambda_L}
\left(
\frac {p_n(x)}{1 - p_n(y)}
\right)^k
=0, 
\quad
k \ge 2.
\eeq
}
%%%
%
%
\begin{proof}
By assumption, 
$f^{-1}(y)$
is well-defined for sufficiently large 
$y$.
We 
then compute
\beq
p_n (x)
&=&
{\bf P}
\Bigl(
\omega (n)
\ge
%\left(
\langle n \rangle^{\alpha}
f^{-1} ( \Gamma_L x )
%\right)
\Bigr)
=
%\\
%
%&=&
\frac {1}{
f
\left(
\langle n \rangle^{\alpha}
f^{-1} ( \Gamma_L x )
\right)
}.
\eeq
The monotonicity of 
$f$
yields 
\beq
\frac {1}{
f
\left(
\langle L \rangle^{\alpha}
f^{-1} ( \Gamma_L x )
\right)
}
\le
p_n (x)
\le
\frac {1}{
\Gamma_L x 
}
\eeq
from which we have 
(0).
(1)
folllows from 
(0)
and Assumption 2.
(2)
follows from 
(1)
and the fact that 
$\lim_{L \to \infty}
\max_{n \in \Lambda_L}
p_n (x)^k
= 0$.
\QED
\end{proof}
We next consider 
the multiplication operator 
$V$ 
and let 
$\{ E^V_j \}_j$
$E^V_1 \ge E^V_2 \ge \cdots$
be the eigenvalues of 
$V$ 
in decreasing order. 
Moreover, let 
\beq
\widetilde{E}^{V} (j)
=
\widetilde{E}^{V}_j
:=
\frac {
f(E^{V}_j)
}
{
\Gamma_L
}, 
\quad
%\sharp = H, V
\eeq
be the rescaling of those. 
Then we show\\

%%%%%
{\bf Lemma 2 }
{\it 
The point process 
$\xi^V_L:=
\sum_{j \ge 1}
\delta_{ \widetilde{E}^V_j }$
whose atoms are 
$\{ \widetilde{E}^V_j \}_j$ 
converges to Poisson 
($\nu$).
}
\begin{proof}
Fix 
$k \in {\bf N}$
and 
let 
$x_1, \cdots, x_k \in {\bf R}$
with 
$0 < x_k < x_{k-1} < \cdots < x_1$
and consider the following intervals: 
\beq
I_k := [x_k, x_{k-1}), 
\cdots, 
I_i := [x_i, x_{i-1}), 
\cdots, 
I_2 := [x_2, x_1), 
\;
I_1 := [x_1, \infty), 
\eeq
Take 
$a_1, a_2, \cdots, a_k \in {\bf Z}_{\ge 0}$, 
let 
$M:=
a_1 + a_2 + \cdots + a_k$ 
and consider the following event. 
\beq
P :=
{\bf P}
\left(
\bigcap_{i=1}^k
\left\{ 
\sharp \{ j \, | \, 
\widetilde{E}^V_j \in I_i \} = a_i
\right\}
\right).
\eeq
It then suffices to show 
%
%%%%%
\beq
{\bf P}
\left(
\bigcap_{i=1}^k
\left\{ 
\sharp \{ j \, | \, 
\widetilde{E}^V_j \in I_i \} = a_i
\right\}
\right)
\stackrel{L \to \infty}{\to}
\frac {1}{a_1! a_2! \cdots a_k !}
\left\{
\prod_{i=2}^k
\left(
\frac {1}{x_i} - \frac {1}{x_{i-1}}
\right)^{a_i}
\right\}
e^{- \frac {1}{x_k}}.
\eeq
%
%%%%%
%
We compute : 
\beq
P
&=&
{\bf P}
\left(
\begin{array}{cl}
\widetilde{E}^V(n) \in I_1, 
&
n = 
1, 2, \cdots, a_1
%n^{(1)}_1, \cdots, n^{(1)}_{a_1} 
\\
\widetilde{E}^V(n) \in I_2, 
&
n = 
a_1 + 1, \cdots, a_1 + a_2
%n^{(2)}_1, \cdots, n^{(2)}_{a_2}
\\
\vdots & \vdots \\
\widetilde{E}^V(n) \in I_i, 
&
n = 
\sum_{j=1}^{i-1} a_{j}+ 1, 
\cdots, 
\sum_{j=1}^{i-1} a_{j}+ a_i
%n^{(i)}_1, \cdots, n^{(i)}_{a_i}
\\
\vdots & \vdots \\
\widetilde{E}^V(n) \in I_k, 
&
n = 
\sum_{j=1}^{k-1} a_{j}+ 1, 
\cdots, 
\sum_{j=1}^{k-1} a_{j}+ a_k
%n^{(k)}_1, \cdots, n^{(k)}_{a_k}
\\
%
%\{ n^{(i)}_j \}_{
%\substack{i=1, \cdots, k \\ j = 1, \cdots, a_i}
%}
%&
%(\subset \Lambda_L)
%: 
%distinct
%
\end{array}
\right)
\\
&=&
{\bf P}
\left(
\begin{array}{cc}
\widetilde{V}(n) \in I_1, 
&
n = 
n^{(1)}_1, \cdots, n^{(1)}_{a_1} 
\\
\widetilde{V}(n) \in I_2, 
&
n = 
n^{(2)}_1, \cdots, n^{(2)}_{a_2}
\\
\vdots & \vdots \\
\widetilde{V}(n) \in I_i, 
&
n = 
n^{(i)}_1, \cdots, n^{(i)}_{a_i}
\\
\vdots & \vdots \\
\widetilde{V}(n) \in I_k, 
&
n = 
n^{(k)}_1, \cdots, n^{(k)}_{a_k}
\\
\widetilde{V}(n) \notin  \bigcup_{i=1}^k I_i, 
&
n \notin 
\{ n^{(i)}_j \}_{
\substack{i=1, \cdots, k \\ j = 1, \cdots, a_i}
}
\\
\{ n^{(i)}_j \}_{
\substack{i=1, \cdots, k \\ j = 1, \cdots, a_i}
}
&
(\subset \Lambda_L)
: 
distinct
\end{array}
\right)
\\
&=&
\frac {1}{a_1! a_2! \cdots a_k !}
\sum_{
\{ n^{(i)}_j \}_{
\substack{i=1, \cdots, k \\ j = 1, \cdots, a_i}
}
}
\prod_{i=1}^k
\prod_{j=1}^{a_i}
\frac {
p_{n^{(i)}_j} (x_i) - p_{n^{(i)}_j} (x_{i-1})
}
{
1 - p_{n^{(i)}_j}(x_k)
}
\cdot
(1 - p_{n^{(i)}_j}(x_k))
\\
&& \qquad
\times
\prod_{n \in \Lambda_L 
\setminus \{ n^{(i)}_j \}_{
\substack{i=1, \cdots, k \\ j = 1, \cdots, a_i}
}
}
(1 - p_n (x_k))
\cdot
1\left(
\{ n^{(i)}_j \}_{
\substack{i=1, \cdots, k \\ j = 1, \cdots, a_i}
}
%\stackrel{i=1, \cdots, k}{j = 1, \cdots, a_i} }
: distinct
\right)
\\
&=&
\frac {1}{a_1! a_2! \cdots a_k !}
\sum_{
\{ n^{(i)}_j \}_{
\substack{i=1, \cdots, k \\ j = 1, \cdots, a_i}
}
}
\prod_{i=1}^k
\prod_{j=1}^{a_i}
\frac {
p_{n^{(i)}_j} (x_i) - p_{n^{(i)}_j} (x_{i-1})
}
{
1 - p_{n^{(i)}_j}(x_k)
}
%\cdot(1 - p_{n^{(i)}_j}(x_M))
\\
&& \qquad
\times
\prod_{n \in \Lambda_L}
(1 - p_n (x_k))\cdot
1\left(
\{ n^{(i)}_j \}_{
\substack{i=1, \cdots, k \\ j = 1, \cdots, a_i}
}
%\stackrel{i=1, \cdots, k}{j = 1, \cdots, a_i} }
: distinct
\right).
\eeq
Here 
we substitute 
\beq
1\left(
\{ n^{(i)}_j \}_{
\substack{i=1, \cdots, k \\ j = 1, \cdots, a_i}
}
%\stackrel{i=1, \cdots, k}{j = 1, \cdots, a_i} }
: distinct
\right)
&=&
1 - 
1 \left(
\exists
(i,j), (i', j')
\;s.t.\;
n^{(i)}_j = n^{(i')}_{j'}
\right)
\eeq
and decompose RHS into two terms, 
\beq
P
&=&
P_1 - P_2.
\eeq
We show that 
$P_1$ 
converges to the right one, while 
$P_2 \to 0$. 
For 
$P_1$, 
\beq
P_1
&=&
\frac {1}{a_1! a_2! \cdots a_k !}
\sum_{
\{ n^{(i)}_j \}_{
\substack{i=1, \cdots, k \\ j = 1, \cdots, a_i}
}
}
\prod_{i=1}^k
\prod_{j=1}^{a_i}
\frac {
p_{n^{(i)}_j} (x_i) - p_{n^{(i)}_j} (x_{i-1})
}
{
1 - p_{n^{(i)}_j}(x_k)
}
%\cdot(1 - p_{n^{(i)}_j}(x_M))
%
\times
\prod_{n \in \Lambda_L}
(1 - p_n (x_k))
\\
&=&
\frac {1}{a_1! a_2! \cdots a_k !}
\prod_{i=1}^k
\prod_{j=1}^{a_i}
\sum_{
n^{(i)}_j \in \Lambda_L
}
\frac {
p_{n^{(i)}_j} (x_i) - p_{n^{(i)}_j} (x_{i-1})
}
{
1 - p_{n^{(i)}_j}(x_k)
}
%\cdot(1 - p_{n^{(i)}_j}(x_M))
%
\times
\prod_{n \in \Lambda_L}
(1 - p_n (x_k))
\\
& \stackrel{L\to \infty}{\to} &
\frac {1}{a_1! a_2! \cdots a_k !}
\prod_{i=1}^k
\left(
\frac {1}{x_i} - \frac {1}{x_{i-1}}
\right)^{a_i}
e^{- \frac {1}{x_k}}
\eeq
where we used Lemma 1 and the following fact.
\beq
\log 
\prod_{n \in \Lambda_L}
(1 - p_n (x))
&=&
\sum_{n \in \Lambda_L}
\log (1 - p_n (x))
\\
&=&
- 
\sum_{n \in \Lambda_L} 
p_n (x)
+
{\cal O}
\left(
\sum_{n \in \Lambda_L}
p_n (x)^2
\right)
\stackrel{L \to \infty}{\to}
- \frac 1x.
\eeq

For 
$P_2$, 
\beq
P_2
&=&
\frac {1}{a_1! a_2! \cdots a_k !}
\sum_{
\{ n^{(i)}_j \}_{
\substack{i=1, \cdots, k \\ j = 1, \cdots, a_i}
}
}
\prod_{i=1}^k
\prod_{j=1}^{a_i}
\frac {
p_{n^{(i)}_j} (x_i) - p_{n^{(i)}_j} (x_{i-1})
}
{
1 - p_{n^{(i)}_j}(x_k)
}
%\cdot(1 - p_{n^{(i)}_j}(x_M))
\\
&& \qquad
\times
\prod_{n \in \Lambda_L}
(1 - p_n (x_k))\cdot
1 \left(
\exists
(i,j), (i', j')
\;s.t.\;
n^{(i)}_j = n^{(i')}_{j'}
\right), 
\eeq
we consider the number
$p_M$ 
of all cases to decompose the set 
$[M] = \{ 1, 2, \cdots, M\}$
of 
$M$ 
elements into subsets.
Explicit value of  
$p_M$ 
is given by the Stirling number of second kind : 
$p_M 
=
\sum_{k=1}^M S(M, k)$, 
but we do not need here. 
Let 
$P'$ 
be the one of 
$p_M$ 
terms which constitute 
$P_2$
and let 
%
%\beq
$
n^{(p)}_q
=
n^{(r)}_s
$
%\eeq
%
be the one of those pairings corresponding to  
$P'$. 
We can then estimate 
\begin{equation}
P'
\le 
\frac {1}{a_1! a_2! \cdots a_k !}
\sum_{ 
\{ n^{(i)}_j \}
}
1
\left(
n^{(p)}_q
=
n^{(r)}_s
\right)
\prod_{i=1}^k 
\prod_{j=1}^{a_i}
p(i,j)
\label{Pprime}
%\quad \cdots (*)
\end{equation}
where we set 
\beq
p(i,j)
:=
\frac {
p_{n^{(i)}_j} (x_i) - p_{n^{(i)}_j} (x_{i-1})
}
{
1 - p_{n^{(i)}_j}(x_k)
}
\eeq
Then we estimate
\beq
&&
RHS\; of \;(\ref{Pprime})
\\
&=&
\frac {1}{a_1! a_2! \cdots a_k !}
\sum_{ 
\{ n^{(i)}_j \}
\setminus 
\{ n^{(p)}_q, n^{(r)}_s \}
}
\sum_{n^{(p)}_q, n^{(r)}_s}
%\sum_{ \{ n^{(i)}_j \} }
1
\left(
n^{(p)}_q
=
n^{(r)}_s
\right)
\\
&&
\qquad\qquad\qquad
\left\{
\prod_{i=1}^k 
\prod_{j=1}^{a_i}
p(i,j)
1
\left(
n_j^{(i)} \ne n^{(p)}_q, n^{(r)}_s
\right)
\right\}
p(p,q) \cdot p(r,s)
\\
&=&
\frac {1}{a_1! a_2! \cdots a_k !}
\sum_{ 
\{ n^{(i)}_j \}
\setminus 
\{ n^{(p)}_q, n^{(r)}_s \}
}
\left\{
\prod_{i=1}^k 
\prod_{j=1}^{a_i}
p(i,j)
1
\left(
n_j^{(i)} \ne n^{(p)}_q, n^{(r)}_s
\right)
\right\}
\\
&& 
\times
\sum_{ n \in \Lambda_L }
\frac {
p_{n} (x_p) - p_{n} (x_{p-1})
}
{
1 - p_{n}(x_k)
} 
\cdot
\frac {
p_{n} (x_r) - p_{n} (x_{r-1})
}
{
1 - p_{n}(x_k)
} 
\\
& \le &
\frac {1}{a_1! a_2! \cdots a_k !}
\prod_{i=1}^k 
\prod_{j=1}^{a_i}
1
\left(
(i,j) \ne (p,q), (r,s)
\right)
\times
\\
&&
\qquad\qquad\qquad
\times
\left(
\sum_{ n^{(i)}_j \in \Lambda_L} 
p(i,j)
\right)
%\\
%
%&& 
%\times
%\left(
\sum_{n \in \Lambda_L}
\left(
\frac {
p_{n} (x_p) - p_{n} (x_{p-1})
}
{
1 - p_{n}(x_k)
} 
\right)^2
%\right)^{1/2}
%
%\left(
%\sum_{n \in \Lambda_L}
%\left(
%\frac {
%p_{n} (x_r) - p_{n} (x_{r-1})
%}
%{
%1 - p_{n}(x_k)
%} 
%\right)^2
%\right)^{1/2}
\\
&=&
o(1).
\eeq
\QED
\end{proof}
Finally, 
we compare 
$\widetilde{E}^H$
and 
$\widetilde{E}^V$
to finish the proof of Theorem 2. 
This is elementary, 
but we mention it for completeness.
\\
%

%　
%
%%%%%
{\bf Lemma 2 }
$\xi_L^V
%\{\widetilde{E}^V_j \}_j 
\stackrel{d}{\to} Poisson (\nu)$
implies
$\xi_L
%\{\widetilde{E}^H_j \}_j
\stackrel{d}{\to} Poisson (\nu)$.
%
%%%%%
%
\begin{proof}
By Assumption 1(1), 
for any 
$\epsilon > 0$
we can find 
$M_{\epsilon} > 0$
such that 
$| f' (x) | \le 
\epsilon f(x)$
for 
$x > M_{\epsilon}$. 
And 
by Assumption 1(2), 
we can find positive constant 
$M'$ 
such 
$\sup_{|x-y| \le 2d} |f(y)| \le C |f(x)|$
for 
$x > M'$.
Set 
$M'_{\epsilon}
:=
\max \{ M_{\epsilon}, M' \}$.
Take any 
$j \in {\bf N}$. 
By 
mean value theorem, 
\begin{equation}
|
\widetilde{E}^H_j
-
\widetilde{E}^V_j
|
=
\left|
\frac {f(E_j^H)}{\Gamma_L}
-
\frac {f(E_j^V)}{\Gamma_L}
\right|
=
\left|
\frac {
f'( E'_j) 
}
{
\Gamma_L
}
(E_j^H - E_j^V)
\right|, 
\quad
E'_j 
\in
[ E_j^H, E_j^V ]
\label{meanvalue}
\end{equation}
Since  
$\xi_L^V
%\{ \widetilde{E}^V_j \} 
\to
Poisson (\nu)$, 
we can suppose 
$\lim_{L \to \infty}
E_j^V= \infty$, 
and since 
$| E_j^H - E_j^V | \le 2d$, 
we can also suppose 
$E^H_j, E_j' \to \infty$, 
under those events to be considered below. 
Hence 
by taking 
$L$
sufficiently large such that 
$E^V_j, E^H_j > M'_{\epsilon}$, 
we have 
\beq
| f'(E_j') | \le 
\epsilon
| f(E_j') |
\le
\epsilon
\sup_{|x - E^{\sharp}_j| \le 2d}
| f(x)|
\le
C 
\epsilon
| f(E^{\sharp}_j) |, 
\quad
\sharp = H, V
\eeq
Plugging 
this equation into 
(\ref{meanvalue})
yields 
\begin{equation}
|
\widetilde{E}^H_j
-
\widetilde{E}^V_j
|
 \le 
C \epsilon
\widetilde{E}^{\sharp}_j
\cdot
(2d).
\label{compare}
\end{equation}
Let 
$I_j = (a_j, b_j)$, 
$j = 1, 2, \cdots, K$, 
$0 < a_j < b_j < \infty$
be the disjoint intervals, and let 
$k_j \in {\bf N} \cup \{ 0 \}$. 
For 
$\delta > 0$
sufficiently small, let 
$I_j^{\pm\delta} := (a_j \mp \delta, b_j \pm \delta)$
be the intervals inflated (resp. shrinked) by 
$\delta$
from 
$I_j$.
Then by (\ref{compare}), 
we have for 
$L$
sufficiently large, 
\begin{eqnarray}
&&
\left\{
\sharp \{ 
\widetilde{E}_i^H \in I
\} 
= k 
\right\}
\\
& \subset &
\left\{
\sharp \{ \widetilde{E}^V_i \in I^{+\delta} \} \ge k
\right\}
\cap
\left\{
\sharp \{ 
\widetilde{E}_i^H \in I
\} 
= k 
\right\}
\nonumber
\\
&\subset&
\left(
\left\{
\sharp \{ \widetilde{E}^V_i \in I^{+\delta} \} = k
\right\}
\right)
\cup 
\left(
\left\{
\sharp \{ \widetilde{E}^V_i \in I^{+\delta} \} \ge k+1
\right\}
\cap
\left\{
\sharp \{ 
\widetilde{E}_i^H \in I
\} 
= k 
\right\}
\right)
\nonumber
\\
&\subset&
\left(
\left\{
\sharp \{ \widetilde{E}^V_i \in I^{+\delta} \} = k
\right\}
\right)
\cup 
\left(
\left\{
\sharp \{ \widetilde{E}^H_i \in I^{+2\delta} \} \ge k+1
\right\}
\cap
\left\{
\sharp \{ 
\widetilde{E}_i^H \in I
\} 
= k 
\right\}
\right)
\nonumber
\\
& \subset &
\left\{
\sharp \{ \widetilde{E}^V_i \in I^{+\delta} \} = k
\right\}
\cup 
\left\{
\sharp \{ \widetilde{E}^H_i \in I^{+2\delta}\setminus I \} \ge 1
\right\}
\label{intermediate}
%\quad\cdots (*)
%\nonumber
\\
& \subset &
\left\{
\sharp \{ \widetilde{E}^V_i \in I^{+\delta} \} = k
\right\}
\cup 
\left\{
\sharp \{ \widetilde{E}^V_i \in I^{+3\delta}\setminus I^{- \delta} \} \ge 1
\right\}
%\label{upperbound}
\nonumber
\end{eqnarray}
which implies
\begin{equation}
\bigcap_{j=1}^K
\left\{
\sharp \{ 
\widetilde{E}_i^H \in I_j
\} 
=
k_j
\right\}
\subset 
\bigcap_{j=1}^K
\left\{
\sharp \{ \widetilde{E}^V_i \in I_j^{+\delta} \} = k_j
\right\}
\cup
\bigcup_{j=1}^K
\left\{
\sharp \{ \widetilde{E}^V_i \in I_j^{+3\delta}\setminus I^{- \delta}_j \} \ge 1
\right\}.
\label{upperbound}
\end{equation}
By
(\ref{intermediate}), 
\beq
\left\{
\sharp
\{ 
\widetilde{E}_i^H \in I
\}
= k
\right\}
\subset
\left\{
\sharp 
\{ 
\widetilde{E}_i^V \in I^{+\delta} 
\} = k
\right\}
\cup
\left\{
\sharp 
\{ 
\widetilde{E}_i^H \in I^{+2 \delta} \setminus I
\}
\ge 1
\right\}.
\eeq
Switching 
$H$
and 
$V$, 
\beq
\left\{
\sharp
\{ 
\widetilde{E}_i^V \in I
\}
= k
\right\}
\subset
\left\{
\sharp 
\{ 
\widetilde{E}_i^H \in I^{+\delta} 
\} = k
\right\}
\cup
\left\{
\sharp 
\{ 
\widetilde{E}_i^V \in I^{+2 \delta} \setminus I
\}
\ge 1
\right\}.
\eeq
Replacing 
$I$
by 
$I^{- \delta}$, 
\beq
\left\{
\sharp
\{ 
\widetilde{E}_i^V \in I^{- \delta}
\}
= k
\right\}
\subset
\left\{
\sharp 
\{ 
\widetilde{E}_i^H \in I 
\} = k
\right\}
\cup
\left\{
\sharp 
\{ 
\widetilde{E}_i^V \in I^{+ \delta} \setminus I^{- \delta}
\}
\ge 1
\right\}.
\eeq
Therefore, we have
\begin{equation}
\bigcap_{j=1}^K
\left\{
\sharp \{ 
\widetilde{E}_i^V \in I^{-\delta}_j
\} 
=
k_j
\right\}
\subset 
\bigcap_{j=1}^K
\left\{
\sharp \{ \widetilde{E}^H_i \in I_j^{} \} = k_j
\right\}
\cup
\bigcup_{j=1}^K
\left\{
\sharp \{ \widetilde{E}^V_i \in I_j^{+\delta}\setminus I^{- \delta}_j \} \ge 1
\right\}.
\label{lowerbound}
\end{equation}
By 
(\ref{upperbound}), (\ref{lowerbound}), 
\beq
{\bf P}
\left(
\bigcap_{j=1}^K
\left\{
\sharp \{ 
\widetilde{E}_i^V \in I^{-\delta}_j
\} 
=
k_j
\right\}
\right)
&-&
\sum_{j=1}^K
{\bf P}
\left(
\left\{
\sharp \{ \widetilde{E}^V_i \in I_j^{+\delta}\setminus I^{- \delta}_j \} \ge 1
\right\}
\right)
\\
&\le&
{\bf P}
\left(
\bigcap_{j=1}^K
\left\{
\sharp \{ \widetilde{E}^H_i \in I_j^{} \} = k_j
\right\}
\right)
\\
\le
{\bf P}
\left(
\bigcap_{j=1}^K
\left\{
\sharp \{ 
\widetilde{E}_i^V \in I^{+\delta}_j
\} 
=
k_j
\right\}
\right)
&+&
\sum_{j=1}^K
{\bf P}
\left(
\left\{
\sharp \{ \widetilde{E}^V_i \in I_j^{+3\delta}\setminus I^{- \delta}_j \} \ge 1
\right\}
\right).
\eeq
Taking 
$\liminf_L$, 
$\limsup_L$
on both sides yields
\beq
{\bf P}
\left(
\bigcap_{j=1}^K
\left\{
\sharp \{ 
X_i \in I^{-\delta}_j
\} 
=
k_j
\right\}
\right)
&-&
\sum_{j=1}^K
{\bf P}
\left(
\left\{
\sharp \{ X_i \in I_j^{+\delta}\setminus I^{- \delta}_j \} \ge 1
\right\}
\right)
\\
\le
\liminf_{L \to \infty}
{\bf P}
\left(
\bigcap_{j=1}^K
\left\{
\sharp \{ \widetilde{E}^H_i \in I_j^{} \} = k_j
\right\}
\right)
&\le&
\limsup_{L \to \infty}
{\bf P}
\left(
\bigcap_{j=1}^K
\left\{
\sharp \{ \widetilde{E}^H_i \in I_j^{} \} = k_j
\right\}
\right)
\\
\le
{\bf P}
\left(
\bigcap_{j=1}^K
\left\{
\sharp \{ 
X_i \in I^{+\delta}_j
\} 
=
k_j
\right\}
\right)
&+&
\sum_{j=1}^K
{\bf P}
\left(
\left\{
\sharp \{ X_i \in I_j^{+3\delta}\setminus I^{- \delta}_j \} \ge 1
\right\}
\right), 
\eeq
where 
$\{ X_j \}_{j=1}^{\infty} \sim Poisson (\nu)$.
Since 
${\bf P}
\left(
\sharp \{ X_i \in I \}
\ge 1
\right)
\stackrel{\nu(I) \to 0}{=}
{\cal O}( \nu (I) )
$, 
we have 
\beq
{\bf P}
\left(
\bigcap_{j=1}^K
\left\{
\sharp \{ 
X_i \in I^{-\delta}_j
\} 
=
k_j
\right\}
\right)
&-&
(Const.)
\sum_{j=1}^K
\nu 
\left(
I_j^{+2\delta}\setminus I^{-2 \delta}_j
\right)
\\
\le
\liminf_{L \to \infty}
{\bf P}
\left(
\bigcap_{j=1}^K
\left\{
\sharp \{ \widetilde{E}^H_i \in I_j^{} \} = k_j
\right\}
\right)
&\le&
\limsup_{L \to \infty}
{\bf P}
\left(
\bigcap_{j=1}^K
\left\{
\sharp \{ \widetilde{E}^H_i \in I_j^{} \} = k_j
\right\}
\right)
\\
\le
{\bf P}
\left(
\bigcap_{j=1}^K
\left\{
\sharp \{ 
X_i \in I^{+\delta}_j
\} 
=
k_j
\right\}
\right)
&+&
(Const.)
\sum_{j=1}^K
\nu 
\left(
I_j^{+3\delta}\setminus I^{- \delta}_j
\right).
\eeq
Taking 
$\delta \to 0$, 
we get the desired conclusion.
\QED
\end{proof}
%

%
%%%%%%%%%%%%%%%%%%%%%%%%%%%%%%%%%%%%%%%%%%%%%%%%%%%%%%%%%%%%
\section{Proof of Theorem 3}
It is easy 
to see that 
$f_{p, k}$
satisfies Assumption 1.
For 
Assumption 2, 
we discuss the cases 
$\alpha p < d$
and
$\alpha p = d$
separately.
In this section we simply write
$f = f_{p, k}$. 
%
%%%%%%%%%%%%%%%%%%%%%%%%%%%%%%%%%%%%
\subsection{
$\alpha p < d$}
%
%
%%%%%%%%%
{\bf Lemma 3 }
Let 
\beq
\Gamma_L := \gamma_{p, k} L^{d - \alpha p}, 
\quad
\gamma_{p, k}
:=
\frac {
C_{d-1}
}
{
d - \alpha p
}
\left(
\frac {d}{d - \alpha p}
\right)^k
\eeq

Then 
$f$
and 
$\Gamma_L$ 
satisfy Assumption 2 : 
\beq
\sum_{n \in \Lambda_L}
p_n (x)
=
\frac 1x
+
o(1), 
\quad
L \to \infty.
\eeq
%
%%%
%
\begin{proof}
It is straightforward to see 
\beq
\sum_{n \in \Lambda_L}
p_n (x)
&=&
\frac {1}{
\Gamma_L x
}
\sum_{n \in \Lambda_L}
\frac {1}{
\langle n \rangle^{\alpha p}
}
\left(
1 + 
\dfrac {\log \langle n \rangle^{\alpha}}
{
\log f^{-1} (\Gamma_Lx)
}
\right)^k
\\
&=&
\frac {1}{
\Gamma_L x
}
\sum_{n \in \Lambda_L}
\frac {1}{
\langle n \rangle^{\alpha p}
}
\sum_{l=0}^k
\left(
\begin{array}{c}
k \\ l
\end{array}
\right)
\left(
\dfrac {\log \langle n \rangle^{\alpha}}
{
\log f^{-1} (\Gamma_Lx)
}
\right)^l
\\
&=:&
\sum_{l=0}^k
A_l
\\
A_l
&:=&
\frac {1}{
\Gamma_L x
}
%\sum_{l=0}^k
%
\left(
\begin{array}{c}
k \\ l
\end{array}
\right)
\sum_{n \in \Lambda_L}
\frac {
(\log \langle n \rangle^{\alpha})^l
}{
\langle n \rangle^{\alpha p}
}
\dfrac {
1
}
{
(\log f^{-1} (\Gamma_Lx))^l
}.
\eeq
We here note that 
\begin{equation}
y 
=
\log f^{-1}(z)
=
\frac 1p
\log z
+
\frac kp
\log \log z
(1 + o(1)), 
\quad
z \to \infty.
\label{asymptotic}
\end{equation}
To see 
(\ref{asymptotic}), 
we first set 
$y:= \log x$, 
and 
$g(y) := y^{-k} e^y$.
Then 
$z = f(x)$
if and only if 
$y = 
p^{-1}
%\frac 1p 
g^{-1} (p^{-k}z)$
so that we study the asymptotic behavior of 
$g^{-1}(z)$ 
as 
$z \to \infty$.
It then 
suffices to integrate the following formula : 
\beq
\frac {dy}{dz}
=
\frac 1z
+
k
\cdot
\frac 1z
\cdot
\frac {1}{\log z}
+
{\cal O}
\left(
\frac {\log \log z}{z (\log z)^2}
%\cdot\frac {\log y}{y}
\right).
\eeq
By 
(\ref{asymptotic}), 
we have 
\beq
\log 
f^{-1} ( \Gamma_L x )
=
\frac 1p
\log \Gamma_L
(1 + o(1)).
\eeq
On the other hand, 
since the leading term of the diverging series is given by the corresponding integral, we have 
\beq
\sum_{n \in \Lambda_L}
\frac {
(\log \langle n \rangle^{\alpha})^l
}{
\langle n \rangle^{\alpha p}
}
&=&
\alpha^l
C_{d-1}
\int_1^L
dx
x^{d - 1 - \alpha p}
( \log x )^l 
(1 + o(1))
\\
&=&
\frac {
\alpha^l
C_{d-1}
}
{
d - \alpha p
}
L^{d - \alpha p}
( \log L )^l 
( 1 + o(1)).
\eeq
Plugging them into the definition of 
$A_l$
yields
\beq
A_l
&=&
\frac {1}{
\Gamma_L x
}
%\sum_{l=0}^k
%
\left(
\begin{array}{c}
k \\ l
\end{array}
\right)
\frac {
(\alpha p)^l
}
{
d - \alpha p
}
C_{d-1}
\frac {
L^{d - \alpha p} 
( \log L )^l 
}
{
(\log \Gamma_L)^l
}
(1 + o(1)), 
\quad
l = 0, 1, \cdots, k.
\eeq
Since 
%
%\beq
$
A_l
\sim
L^{d - \alpha p}
\dfrac {
(\log L)^l
}
{
\Gamma_L ( \log \Gamma_L )^l
}
$, 
we would like to choose 
$\Gamma_L$
such that 
%
%\beq
$
L^{d - \alpha p}
(\log L)^l
\sim
\Gamma_L ( \log \Gamma_L )^l 
$.
%\eeq
%
Thus 
we take a constant
$\gamma$
to be fixed later and set 
$
\Gamma_L
:=
\gamma 
L^{d - \alpha p}
$.
Then 
\beq
A_l
&=&
\frac 1x
\left(
\begin{array}{c}
k \\ l
\end{array}
\right)
\frac {C_{d-1}}
{
d - \alpha p
}
\frac {1}{\gamma}
\left(
\frac {\alpha p}
{
d - \alpha p
}
\right)^l
\cdot
\frac {1}{
\left(
1 + 
\dfrac {\log \gamma}
{
(d - \alpha p) \log L
}
\right)^l
}
(1 + o(1))
\\
\sum_{n \in \Lambda_L}
p_n (x)
&=&
\sum_{l=0}^k 
A_l
=
\frac 1x
\frac {C_{d-1}}
{
d - \alpha p
}
\frac {1}{\gamma}
\left(
\frac {d}{d - \alpha p}
\right)^k
(1 + o(1)).
\eeq
Therefore 
choosing 
$\gamma = \gamma_{p, k}$, 
we have Assumption 2.
\QED
\end{proof}
%
%
%
%%%%%%%%%%%%%%%%%%%%%%%%%%%%%%%%%%%%
\subsection{
$\alpha p = d$, 
$q = -k$, 
$k \in {\bf N}  \cup \{ 0 \}$
}
%

%
%%%%%%%%%
{\bf Lemma 4 }
Let 
$\alpha p = d$, 
$q = -k$, 
$k \in {\bf N} \cup \{ 0 \}$.
Let 
$\Gamma_L$ 
satify the following equation.
\beq
\Gamma (\log \Gamma)^k
=
\gamma
(\log L)^{k+1}, 
\quad
\gamma 
:=
\frac {C_{d-1}}{k+1}
\cdot
p^k.
\eeq
Then 
$f$
and 
$\Gamma_L$ 
satisfy Assumption 2.  
\begin{proof}
As 
in the proof of Lemma 3, we have 
\beq
\sum_{n \in \Lambda_L}
p_n (x)
&=&
\frac {1}{
\Gamma_L x
}
\sum_{n \in \Lambda_L}
\frac {1}{
\langle n \rangle^{\alpha p}
}
\left(
1 + 
\dfrac {\log \langle n \rangle^{\alpha}}
{
\log f^{-1} (\Gamma_Lx)
}
\right)^k
=:
\sum_{l=0}^k
A_l
\\
A_l
&:=&
\frac {1}{
\Gamma_L x
}
%\sum_{l=0}^k
%
\left(
\begin{array}{c}
k \\ l
\end{array}
\right)
\sum_{n \in \Lambda_L}
\frac {
(\log \langle n \rangle^{\alpha})^l
}{
\langle n \rangle^{\alpha p}
}
\dfrac {
1
}
{
(\log f^{-1} (\Gamma_Lx))^l
}.
\eeq
Plugging 
\beq
\sum_{n \in \Lambda_L}
\frac {
(\log \langle n \rangle^{\alpha})^l
}{
\langle n \rangle^{\alpha p}
}
&=&
\frac {C_{d-1}}{l+1}
(\log L)^{l+1}
(1 + o(1))
, 
\\
(\log f^{-1}(\Gamma_L x))^l
&=&
\frac {1}{p^l}
(\log \Gamma_L)^l (1+o(1))
\eeq
into 
$A_l$
yields
\beq
A_l
&=&
\frac {1}{
\Gamma_L x
}
\cdot
\left(
\begin{array}{c}
k \\ l
\end{array}
\right)
\cdot
\frac {C_{d-1}}{l+1}
(\log L)^{l+1}
\cdot
\frac {1}
{
\dfrac {1}{p^l}
(\log \Gamma_L)^l
}
(1+o(1))
\sim 
\frac {
(\log L)^{l+1}
}
{
\Gamma_L (\log \Gamma_L)^l
}.
\eeq
Since 
$A_k$ 
has major contribution, we will choose 
$\Gamma_L$ 
such that 
\beq
(\log L)^{k+1}
\sim
\Gamma (\log \Gamma)^k
\eeq
which would imply
\beq
\frac {
(\log L)^{l+1}
}
{
\Gamma_L (\log \Gamma_L)^l
}
&=&
\left(
\frac {
\log \Gamma
}
{
\log L
}
\right)^{k-l}
\le
\left(
\frac {
\log \log L
}
{
\log L
}
\right)^{k-l}
=
o(1), 
\quad
l = 0, 1, \cdots, k-1
\eeq
and thus, 
\beq
A_k 
&=&
{\cal O}(1), 
\quad
A_0, A_1, \cdots, A_{k-1} = o(1).
\eeq
Therefore let 
$\Gamma_L$ 
satisfy
\beq
\Gamma (\log \Gamma)^k
=
\gamma
(\log L)^{k+1}.
\eeq
Then 
\beq
\sum_{n \in \Lambda_L}
p_n (x)
=
A_k + o(1)
=
\frac {1}{x}
\cdot
\frac {
C_{d-1}
}
{
k+1
}
\cdot
p^k
\cdot
\frac {1}{\gamma}
\eeq
which leads us to the conclusion.
\QED
\end{proof}
%
%
%%%%%%%%%%%%%%%%%%%%%%%%%%%%%%%%%%%%%%%%%%%%%%%%%%%%%%%%%%%%
\section{Proof of Theorem 4}
%
%
%%%%%%%%%%%%%%%%%%%%%%%%%%%%%%%%%%%%
\subsection{Reduction to the study of 
$E_1^V$}
Let
\beq
A_L (x)
&:=&
{\bf P}
\left(
E^V_1 (L) \le x
\right).
\eeq
Then 
the following the lemma reduces our problem to a simpler one.\\
{\bf Lemma 5}\\
{\it 
Suppose that 
we can find positive constants
$C_j$, $j=1,2$
such that for large 
$x> 0$, 
\begin{equation}
1 - C_1 e^{- x^{\delta}}
\le
\lim_{L \to \infty}
A_L (x)
\le
\exp
\left[
- C_2
x^{ - d / \alpha }
e^{- 2 D_{\alpha, \delta} x^{\delta}}
\right].
\label{asterisque}
%\quad \cdots (*)
\end{equation}
Then 
we have Theorem 4(2).
}
\begin{proof}
Let 
$\Omega_L (x)
:=
\left\{
\omega_n 
\le 
\langle n \rangle^{\alpha} x, 
\;
\forall n \in \Lambda_L
\right\}$.
Then 
we have 
$\{ 
E_1^V (L) \le x 
\} = 
\Omega_L (x)$
and since the sequence of events 
$\{ 
\Omega_L (x)
\}_{L \ge 1}$
is monotonically decreasing, we have
\beq
\lim_{L \to \infty}
A_L (x)
&=&
%\lim_{L \to \infty}
{\bf P}
\left(
\bigcap_{L \ge 1}
\Omega_L
\right)
=
{\bf P}
\left(
\bigcap_{L \ge 1}
\left\{
E_1^V (L) \le x
\right\}
\right).
\eeq
Suppose we have 
(\ref{asterisque}).
Then 
the inclusion 
$\{ E_1^V (L) \le x - 2d \}
\subset
\{ E_1^H (L) \le x  \}
\subset 
\{ E_1^V (L) \le x + 2d \}$
and some elementary manupulations of inequalities yield the statement of Theorem 4(2).
\QED
\end{proof}
%
%
%%%%%%%%%%%%%%%%%%%%%%%%%%%%%%%%%%%%
\subsection{Proof of Theorem 4(2) : lower bound}
In this subsection
we prove the first inequality in 
(\ref{asterisque}).
By definition we have 
%
%\beq
$
A_L (x)
=
\prod_{n \in \Lambda_L}
{\bf P}
\left(
\omega_n \le
\langle n \rangle^{\alpha} x
\right)
=
\prod_{n \in \Lambda_L}
\left(
1 - e^{- \langle n \rangle^{\alpha \delta} x^{\delta}}
\right)
$, 
and thus  
\begin{equation}
B(x)
:=
\lim_{L \to \infty}
\log A_L (x)
=
\sum_{n \in {\bf Z}^d}
\log
\left(
1 - e^{- \langle n \rangle^{\alpha \delta} x^{\delta}}
\right)
=
-
\sum_{k \ge 1}
\frac 1k
\sum_{n \in {\bf Z}^d}
e^{- k\langle n \rangle^{\alpha \delta} x^{\delta}}.
\label{star}
%\quad\cdots (*)
\end{equation}
In what follows, 
we shall denote by 
$C$ 
the universal constants in those estimates below which may differ from line to line.
Since the maximum point 
$y_{max}$
of the function 
$g(y) = y^{d-1} e^{- k x^{\delta} y^{\alpha \delta}}$ 
is in the order of 
$x^{- 1/ \alpha}$, 
we can bound the series by the integral : 
\beq
\sum_{n \in {\bf Z}^d}
e^{- k\langle n \rangle^{\alpha \delta} x^{\delta}}
&=&
\sum_{n=0}^L
\sum_{ |m|_{\infty} = n}
e^{- k\langle n \rangle^{\alpha \delta} x^{\delta}}
\le
C
\int_1^{\infty}
y^{d-1}
e^{- k y^{\alpha \delta} x^{\delta}} dy
+
C 
e^{- k x^{\delta}}
\\
&=&
C
\frac {1}{\alpha \delta}
\frac {1}{k^{\frac {d}{\alpha \delta}}}
\frac {1}{x^{\frac {d}{\alpha}}}
\int_{k x^{\delta}}^{\infty}
z^{\frac {d}{\alpha \delta} - 1}
e^{-z} dz
+
C 
e^{- k x^{\delta}}.
\eeq
Plugging it into 
(\ref{star})
yields
\begin{equation}
B(x)
\ge
-
C
\frac {1}{\alpha \delta}
\frac {1}{x^{\frac {d}{\alpha}}}
\sum_{k \ge 1}
\int_{k x^{\delta}}^{\infty}
z^{\frac {d}{\alpha \delta} - 1}
e^{-z} dz
\frac {1}{k^{1+\frac {d}{\alpha \delta}}}
+
C
\log 
\left(
1 - e^{-x^{\delta}}
\right).
\label{starstar}
%\quad\cdots (**)
\end{equation}
To estimate the integral in 
(\ref{starstar}), 
we set 
$M :=
\left[
\dfrac {d}{\alpha \delta}
\right]$.
Then 
we have 
\begin{eqnarray}
\int_{k x^{\delta}}^{\infty}
z^{\frac {d}{\alpha \delta} - 1}
e^{-z} dz
&\le&
\int_{k x^{\delta}}^{\infty}
z^{M}
e^{-z} dz
=
e^{- k x^{\delta}}
\sum_{n=0}^M
\frac {M!}{n!}
(k x^{\delta})^{n}
\nonumber
\\
&\le&
(M+1)!
e^{- k x^{\delta}}
x^{\delta M}
k^M.
%\nonumber
\label{sharp}
\end{eqnarray}
In what follows 
we assume 
$M \ge 1$ ; 
the argument for 
$M=0$
is similar.
Plugging 
(\ref{sharp})
into 
(\ref{starstar}) 
and using 
$M \le \dfrac {d}{\alpha \delta}$, 
we have 
\beq
B(x)
&:=&
\lim_{L \to \infty} \log A_L (x)
\\
& \ge &
-
\frac {C}{\alpha \delta}
\frac {1}{x^{\frac {d}{\alpha}}}
\sum_{k \ge 1}
\sum_{k=1}^{\infty}
\frac {1}{
k^{
\frac {d}{\alpha \delta}+1
}
}
(M+1)!
e^{- k x^{\delta}}
x^{\delta M}
k^M
+
C
\log 
\left(
1 - e^{-x^{\delta}}
\right)
\\
&=&
-
\frac {C(M+1)!}{\alpha \delta}
\frac {1}{x^{\frac {d}{\alpha}- \delta M}}
\sum_{k \ge 1}
\frac {e^{- k x^{\delta}}}
{k^{\frac {d}{\alpha \delta}+1-M}}
+
C
\log 
\left(
1 - e^{-x^{\delta}}
\right)
\\
& \ge &
-
\frac {C(M+1)!}{\alpha \delta}
%\frac {1}{x^{\frac {d}{\alpha}- \delta M}}
\sum_{k \ge 1}
\frac {e^{- k x^{\delta}}}
{k^{\frac {d}{\alpha \delta}+1-M}}
+
C
\log 
\left(
1 - e^{-x^{\delta}}
\right)
\\
& \ge &
C_0
\log 
\left(
1 - e^{-x^{\delta}}
\right)
\eeq
where 
$C_0
:=
\frac {CM!}{\alpha \delta}
+
C$.
Therefore 
\beq
B(x) = 
\lim_{L \to \infty}
A_L (x)
\ge
\left(
1 - e^{-  x^{\delta}}
\right)^{
C_0
}
\ge
1 - C_0 e^{- x^{\delta}}
.
\eeq
Here 
we use an inequality 
$
(1 - \theta)^{\rho}
\ge
1 - \rho \theta
$
for 
$0 < \theta \ll 1$, 
$\rho \ge 1$
and obtain the lower bound in Lemma 5.
%
%
%%%%%%%%%%%%%%%%%%%%%%%%%%%%%%%%%%%%
\subsection{Proof of Theorem 4(2) : upper bound}
In this subsection
we prove the second inequality in 
(\ref{asterisque}).
Starting from 
(\ref{star}), 
we aim to have a lower bound of the following series 
\beq
S_L (k)
&:=&
\sum_{n \in \Lambda_L}
e^{- k\langle n \rangle^{\alpha \delta} x^{\delta}}.
\eeq
Then
as in the previous subsection 
we have 
\beq
S_L (k)
\ge
C
\int_1^L
y^{d-1}
e^{- k x^{\delta} \langle y \rangle^{\alpha\delta}}
dy.
\eeq
We note 
$\langle y \rangle^{\alpha \delta}
\le
D_{\alpha, \delta} 
(1 + y^{\alpha \delta})$
so that 
\beq
S_L (k)
& \ge &
C
e^{ - D_{\alpha, \delta} k x^{\delta} }
\int_1^L
y^{d-1} 
e^{- D_{\alpha, \delta} k x^{\delta} y^{\alpha \delta}
}
dy
=
%\\
%
%&=&
C
\frac {1}{\alpha\delta}
\left(
\frac {1}{
D_{\alpha, \delta} k 
}
\right)^{d / \alpha \delta}
\frac {
e^{ - D_{\alpha, \delta} k x^{\delta} }
}
{
x^{d / \alpha}
}
J_L (k)
\\
J_L (k)
&:=&
\int_{ D_{\alpha, \delta} k x^{\delta} }^{ D_{\alpha, \delta} k x^{\delta} L^{\alpha \delta} }
z^{
\frac {d}{\alpha \delta} - 1 
}
e^{-z}
dz.
\eeq
Here we recall 
$M
:=
\left[
\dfrac {d}{\alpha \delta}
\right]$
and assume 
$M \ge 1$
; 
for 
$M=0$, 
the argument becomes simpler. 
For large 
$x$, 
we have 
$D_{\alpha, \delta} k x^{\delta}
\ge
D_{\alpha, \delta}  x^{\delta}
> 1$
so that 
$z^{
\frac {d}{\alpha \delta} - 1
}
\ge
z^{M-1}$.
Thus 
\beq
J_L (k)
& \ge &
\int_{ D_{\alpha, \delta} k x^{\delta} }^{ D_{\alpha, \delta} k x^{\delta} L^{\alpha \delta} }
z^{
M - 1 
}
e^{-z}
dz
\stackrel{L \to \infty}{\to}
e^{-D_{\alpha, \delta} k x^{\delta}}
\sum_{n=0}^{M-1}
\frac{(M-1)!}{n!}
(D_{\alpha, \delta} k x^{\delta})^{n}.
\eeq
Plugging them into 
(\ref{star})
yields
\beq
&&
\left| 
\lim_{L \to \infty} \log A_L (x) 
\right|
\\
&\ge&
\sum_{k \ge 1}
\frac 1k
\lim_{L \to \infty}
S_L (k)
\\
&\ge &
\frac {C}{\alpha\delta}
\frac {1}{
(
D_{\alpha, \delta} 
)^{d / \alpha \delta}
}
\frac {1}
{
x^{d / \alpha}
}
\sum_{k \ge 1}
\frac {1}{
k^{1 + \frac {d}{\alpha \delta}}
}
e^{- 2 D_{\alpha, \delta} k x^{\delta}}
\sum_{n=0}^{M-1}
\frac{(M-1)!}{n!}
(D_{\alpha, \delta} k x^{\delta})^{n}.
\eeq
To have 
a bound of simple form, we pick up 
$k=1$, 
%term and 
$n=0$
term only and 
\beq
\left| 
\lim_{L \to \infty} \log A_L (x) 
\right|
& \ge &
\frac {C}{\alpha\delta}
\frac {1}{
(
D_{\alpha, \delta} 
)^{d / \alpha \delta}
}
\frac {1}
{
x^{d / \alpha}
}
e^{- 2 D_{\alpha, \delta}  x^{\delta}}
\eeq
yielding the second inequality in Lemma 5.

\vspace*{1em}

{\bf Acknowledgement }
The authors 
would like to thank the referee for pointing out an error in the previous version of this manuscript.
This work is partially supported by 
JSPS KAKENHI Grant 
Number 20K03659(F.N.).

%
%%%%% REFERENCES %%%%%%%%%%%%%%%%%%%%%
%
%\small

%

\end{document}